\documentclass{epl}
\shorttitle{Solid State Analog for ...}
\title{ Solid State Analog for  He-McKellar-Wilkens Quantum  Phase}
\author{C. A. de Lima Ribeiro\inst{1}\and  C. Furtado\inst{2}\thanks{E-mail: \email{furtado@fisica.ufpb.br}} \and F. Moraes\inst{3}}
\institute{
  \inst{1} Departamento de F\'{\i}sica,\\
            Universidade Estadual de Feira de Santana\\
              BR116-Norte, Km 3, 44.031-460 \\
               Feira de Santana, BA, Brazil\\
  \inst{2} Departamento de F\'{\i}sica,\\
           Universidade Federal da Para\'{\i}ba\\
           Caixa postal 5008, 58051-970\\
            Jo\~ao Pessoa,PB, Brazil\\
  \inst{3}  Laborat\'{o}rio de F\'{\i}sica Te\'{o}rica e Computacional\\
               Departamento de F\'{\i}sica\\
               Universidade Federal de Pernambuco\\
               50670-901 Recife, PE, Brazil
}
\begin{document}
\maketitle
\begin{abstract}
  In this letter we investigate the quantum dynamics of  a quasiparticle in the presence of a 
charged screw dislocation submitted to a uniform magnetic field. Analysing the quantum scattering  
for this quasiparticle we observed the appearance of a topological quantum phase in the solution 
and demonstrate that this phenomenon is the solid state analog of the He-McKeller-Wilkens effect.
\end{abstract}

Aharonov and Bohm(AB)\cite{aha} have demonstrated that a quantum charge circulating around a
magnetic flux tube acquires a quantum topological phase. This effect has been found experimentally 
by Shambers\cite{sham,pes}. Aharonov and Casher\cite{cas} showed that a particle with a magnetic 
moment moving in a electric field accumulates a quantum phase, the Aharonov-Casher(AC) phase. 
This phenomenon has been observed with a neutron interferometer\cite{cim} and in a neutral atomic 
Ramsey interferometer\cite{san}. He and Mckellar\cite{mac}, and Wilkens\cite{wil}  independently, 
have predicted the existence of a quantum phase that a electric dipole acquires, while circulating 
around, and parallel to, a line of magnetic monopoles. This phase has been denominated in the 
literature the He-Mckellar-Wilkens(HMW) phase. It is the Maxwell dual of the AC phase. A practical 
experimental configuration  to test the HMW phase was proposed by Wei, Han and Wei\cite{wei}. 
In their proposal the electric field of a charged wire polarizes a neutral atom and  a uniform 
magnetic field is applied parallel to the wire. In this field configuration, the neutral  atom 
acquire a HMW phase. In a recent article Anandan\cite{ana} has presented a unified and fully 
relativistic treatment of the interaction of the electric and magnetic dipoles moments of a 
particle with the electromagnetic field. Dowling, Williams and Franson\cite{fra} have proposed a 
unified description of all three phenomena(AB, AC and HMW phases) discussing the Maxwell 
electromagnetic duality relation between the three quantum phases. In their duality analysis they 
predict a fourth phenomenon, which is the dual of the Aharonov-Bohm effect.

Topological defects in space-time can be characterized by  metrics
with zero Riemann-Christoffel curvature tensor everywhere, except on the
defects; i.e., by conic-type curvature singularities~\cite{Sta}. Examples of 
such topological defects are the domain wall~\cite{Vil1}, the cosmic 
string~\cite{Vil1,His} and the global monopole~\cite{Bar}. Cosmic
strings provide a bridge between the physics in microscopic and macroscopic
scales. As examples of topological defects in condensed matter physics  we can
mention  vortices in superconductors or superfluids\cite{Vol}, domain walls in
magnetic  materials, solitons in quasi-one-dimensional polymers,
dislocations and disclinations in solids\cite{Kat} or liquid crystals\cite{lubenski}.

The change of topology of a medium introduced by a linear defect such as   a
disclination or a dislocation in the elastic solid or a cosmic string  in the
cosmos, exerts strong effect on the physical properties of the medium.  The  theory of defects 
in solids is viewed as the analogous of three-dimensional gravity in the approach of Katanaev 
and Volovich\cite{Kat}. In this formalism the boundary conditions imposed  by defects in elastic 
media are accounted for by non-Euclidian metrics.  The theory,  in the  continuum limit, describes 
the solid by a  Riemann-Cartan manifold where  curvature  and torsion   are  associated to 
disclinations and dislocations, respectively, in the medium. The Burgers vector of a dislocation 
is associated to  torsion,  and the Frank angle of a disclination to  curvature. In this theory, 
the elastic deformations introduced in the medium by defects are incorporated in the metric of the 
manifold. The quantum  and classical problems in the Riemann-Cartan manifold 
representing a crystal  with a topological defect  have  been extensively 
analyzed  in  recent years~\cite{furtado,fur2}. Also worth mentioning, two decades ago, Kawamura 
\cite{kaw} and Bausch and coworkers\cite{bau} investigated the scattering  of a single  particle in
dislocated media by a different approach and demonstrated  that the equation that governs the 
scattering  is of Aharonov-Bohm type.

 Linear defects in crystals are produced by breaking and reconstructing chemical bonds. The stress and strain provoked by the creation of these defects cause some bonds in the dislocation core to 
remain broken (dangling bonds). This is the main reason for the electrical activity of dislocations 
in semiconductors. This fact is responsible for dislocations becoming eletrically charged. In 
general, this charge is screened by a cylindrical space charge of opposite sign, formed by ionized 
donors or acceptors.  The potential produced by the 
dangling bonds is modeled in this letter with a linear distribution of charge. In a recent work 
Figielski {\it et al}\cite{fig} have found the solid state analog of the AB effect in 
semiconductors with dislocations. They have demonstrated the occurrence of AB interference for  
particles moving around charged dislocations in a semiconductor, which gives rise to 
magnetoconductance oscillations in the macroscopic sample. Related to this,the existence of a
geometrical phase in screw dislocations has been predicted by two of us in recent article\cite{fur}.

In this letter we use the geometric theory of defects in solids to describe the possibility of 
acquisition of the HMW topological  phase by an electric dipole in the presence of a charged 
screw dislocation. Our system is formed by the metric
\begin{equation}
ds^{2}=  d\rho^{2} + \rho^{2}d\phi^{2}+\left[dz+\left(\frac{b}{2\pi}\right)d\phi\right]^{2}. 
\label{metric}
\end{equation} 
Here  $\beta \equiv b/ 2\pi$, where $b$ is the strengths of the Burgers' vector. We consider that 
the sample is placed in the presence of a uniform magnetic field along the dislocation axis 
(z-direction), with strength $\vec{B}=B {\bf e_{z}}$. We also have the electric field  
produced by the dangling bonds in the screw dislocation, which will have a radial 
component $E=\frac{\kappa}{2\pi\rho}{\bf e_{\rho}}$,with  $\kappa = q/\epsilon d$, where $q$ is 
the total charge, $d$ is the height of the defect and $\epsilon$ is the medium permissivity. 
Such configuration is known as the Read cylinder\cite{read}.

We will consider a Hamiltonian for a neutral particle with an electrical dipole moment, which includes the topological 
information on the medium in its kinetic part derived from the Laplace-Beltrami operator for metric (\ref{metric}). It also 
includes an electric field produced by a charge distribution on the defect. This system is placed in a region with a 
uniform magnetic field along the $z$ direction.

In this context  we will find the Schr\"odinger equation from the Hamiltonian of the quaseparticle with mass in this configuration given by, $m^{*} \equiv M+\alpha B^{2}$, where $M$ is the effective mass of quaseparticle and $\alpha$ is the dielectric polarizability of quaseparticle. The Hamiltonian is described by
\begin{equation}
H= \frac{1}{2m^{*}}\left[\vec{P} + \alpha (\vec{E} \times  \vec{B})\right]^{2}-\frac{1}{2}\alpha E^{2}.
\end{equation}
Observe that the potential which depends on $E^{2}$, represents a local influence to the wave function. We are 
interested in to study the asymptotic states for the atom dynamics. Then we will not consider this term, because to 
us its effect is local \cite{wei}. It contributes with an atractive potential which does not cause an Aharonov-Bohm-like 
scattering \cite{denschlag97}. A complete study of this problem should include an atractive potential of this kind plus 
a repulsive potential originated from deformation potential theory \cite{bausch2}. This will be done in a more detailed 
forthcoming publication. 
Using the Laplace-Beltrami operator, corresponding to the metric (\ref{metric}), as the kinetic energy operator for 
the neutral quasiparticle (which could be an exciton for example) we can write  the time-independent 
Schr\"odinger equation as
\begin{eqnarray}
\label{sch}
-\frac{\hbar^{2}}{2m^{*}}\left[\partial_{\rho\rho}+\frac{1}{\rho}\partial_{\rho}+\partial_{zz}+\frac{1}{\rho^{2}}\left[\left(\partial_{\phi}-\beta\partial_{z}\right)+i\xi \right]^{2}\right]\Psi(\rho,\phi,z)=\nonumber\\E\Psi(\rho,\phi,z),
\end{eqnarray}
where 
\begin{equation}
\xi \equiv \frac{\alpha \kappa B}{2\pi\hbar}.
\end{equation}
>From (\ref{sch}), the Burgers vector couples with the parameter $\xi$ that depends on  electromagnetic field. This coupling produces a combined effect of the torsion produced by defects with the electromagnetic field. 

We solve this equation using the {\it Ansatz}
\begin{equation}
\Psi(\rho,\phi,z)=\Omega(\rho,\phi)\exp[ikz].
\end{equation}
Substituting the Ansatz above in Eq. (\ref{sch}) we obtain
\begin{eqnarray}
-\frac{\hbar^{2}}{2m^{*}}\left[\partial_{\rho\rho}+\frac{1}{\rho}\partial_{\rho}+\frac{1}{\rho^{2}}\left(\partial_{\phi}+i\nu\right)\right]\Omega(\rho,\phi)=\nonumber\\
\left(E-\frac{k^{2}\hbar^{2}}{2m^{*}}\right)\Omega(\rho,\phi), \label{eff}
\end{eqnarray}
where the constant $\nu$ is defined by
\begin{equation}
\nu \equiv +\frac{\alpha\kappa B}{2\pi \hbar}- \frac{b}{2\pi}k,
\end{equation}
and
\begin{equation}
\tilde{E}=E-\frac{k^{2}\hbar^{2}}{2m^{*}},
\end{equation}

We see (\ref{eff}) as an effective Schr\"odinger equation where the effective Hamiltonian describes 
a {\it charged particle} submmited to the {\it effective vector  potential},
\begin{eqnarray*}
  A_{\phi}= \frac{1}{e}(+\frac{\alpha\kappa B}{2\pi \hbar}- \frac{b}{2\pi}k).
\end{eqnarray*}
This kind of potential produces a quantum topological  phase that is given by
\begin{eqnarray}
  \label{pha}
  \Psi_{eHMW}=\frac{e}{\hbar}\oint \vec{A}\dot d \vec{s}=\frac{\alpha\kappa B}{\hbar^{2}} - \frac{bk}{\hbar}.
\end{eqnarray}
Note that this phase produces a combined effect of the torsion produced by the defect (represented 
by the Burgers vector $b$) and the He-Mckellar-Wilkens phase (represented by first term of 
(\ref{pha})). This phase is the solid state analog of the Wei-Han-Wei phase, that is a particular 
case of the He-McKellar-Wilkens phase. 

Now we analyze the scattering of this quasiparticle in the presence of this defect.
Using the Dirac factor method \cite{dirac31} we describe   $\Omega(\rho, \phi)$ as a function of the particle 
wavefunction in the absence of the defect. We seek the solution of (\ref{eff}) in the form below
\begin{equation}
\label{dir}
\Omega(\rho,\phi)=\Lambda(\rho,\phi)\exp\left(-i\int_{0}^{\phi} \nu d\phi'\right),
\end{equation}
 where $\Lambda(\rho,\phi)$ is the solution when we do not have any defect in the medium. Doing 
the following transformations
\begin{equation}
\Lambda(\rho,\phi)=R(\rho)\exp(i\ell\phi)
\end{equation}
and
\begin{equation}  
\tilde{k}^{2}=\frac{2\tilde{E}}{\hbar^{2}},
\end{equation}
we obtain the radial equation,
\begin{equation}
\frac{d^{2}R(\rho)}{d\rho^{2}}+\frac{1}{\rho}\frac{dR(\rho)}{d\rho}+\left(\tilde{k}^{2} - \frac{\ell^{2}}{\rho^{2}}\right)R(\rho)=0.
\end{equation}
The solution to this equation can be described by Bessel and Neumann functions:
\begin{eqnarray}
\Psi(\rho,\phi,z)=\left[ a_{\ell}J_{|\ell |}(\tilde{k}\rho) + b_{\ell}N_{|\ell |}(\tilde{k}\rho)
\right]\exp\left\{i\left[\left(\ell\phi+kz\right)-\int_{0}^{\phi} \nu d\phi'\right]\right\},
\end{eqnarray}
where $a_{\ell}$ and $b_{\ell}$ are constants.
The phase of the  wavefunction (which keeps all information about the topological defect) is 
present in the factor $\nu$, as well as the intensity of the electrical and magnetic fields. Then, 
in order to learn about these phases we need to study the scattering of the quasiparticle by a 
screw dislocation. Some authors already studied a neutral atom scattering by a charged 
line\cite{denschlag97, leonhardt98,audretsch98}. We use the Dirac phase factor  to 
avoid possible problems with a multivalued function.
We consider that the incident wave has the form 
\begin{equation}
\label{inc}
\Psi_{in}(\rho,\phi)=\exp[-i\tilde{k}\rho\cos\phi-i\nu\phi],
\end{equation}
when we follow a loop around the defect, the wavefunction changes by a factor of $\exp(i 2\pi\nu)$. 
We remove this multivalued problem by the {\it whirling wave} procedure\cite{dirac31}.The 
multivaluedness is now explicit in (\ref{inc}), because $\phi$ changes 
by a factor $\exp(2\pi \imath \nu)$ along a closed path around
the defect. The procedure to be adopted consists in decomposing $\Psi_{0}$ into 
an infinite number of components ({\it{whirling waves}}), which are multivalued
and then apply (\ref{dir}) to each whirling wave to get the exact 
single-valued wavefunction. The decomposition of $\Lambda(\rho,\phi)$ in partial 
waves is given by 
\begin{eqnarray}
  \label{eq:561}
\exp(-i\tilde{k}\rho\cos\phi - i\nu\phi) = \sum_{\ell = - \infty}^{+\infty} (-i)^{|\ell |}J_{ |\ell |}(\tilde{k}\rho)\exp(i\phi).
\end{eqnarray}
Using the Poisson summation formula\cite{lighthill58} and performing some manipulations we obtain 
that the wave function can be rewritten in the following form:

\begin{equation}
\Psi(\rho,\phi)= \sum_{m= - \infty}^{+\infty} S_{m}(\rho,\phi),
\end{equation}
where $S_{m}(\rho,\phi)$ is described by
\begin{eqnarray}
S_{m}(\rho,\phi)=\int_{-\infty}^{+\infty} \exp(-i\frac{\pi}{2}|\lambda|) J_{|\lambda |} (k\rho) \exp[i\lambda(\phi+2\pi m)]d\lambda, 
\end{eqnarray}
which is not a single-valued function. But, note that $S_{m}(\rho, \phi+2\pi) = S_{m}(\rho,\phi)$. 
We apply the Dirac phase factor method again and the phase assumes the value  $(\phi+2\pi m)$. The 
sum over m results in,
\begin{equation}
\Psi(\rho,\phi)=  \sum_{m= - \infty}^{+\infty} S_{m}(\rho,\phi) \exp[-i\nu(\phi+2\pi m)],
\end{equation}
and the wave function is given by 
\begin{eqnarray}
\Psi(\rho,\phi)=  \sum_{m= - \infty}^{+\infty}\int_{-\infty}^{+\infty} \exp(-i\frac{\pi}{2}|\lambda|) J_{|\lambda |} (k\rho) \exp[i(\lambda-\nu)(\phi+2\pi m)]d\lambda.
\end{eqnarray}
 
Making the inverse transformation\cite{lighthill58}, we will consider  $|\ell - \nu|$ as a new 
variable
\begin{equation}
\Psi(\rho,\phi)=  \sum_{\ell = - \infty}^{+\infty} (-i)^{|\ell -\nu |}J_{| \ell-\nu |}\exp(i\ell 
\phi).
\end{equation}
Using the asymptotic approximation to the Bessel function we obtain
\begin{eqnarray}
\Psi(\rho,\phi)=  \sum_{\ell = - \infty}^{+\infty} \exp\left(-i\frac{\pi}{2}\left|\ell-\nu\right|\right)\sqrt{\frac{2}{\pi k \rho}}\cos\left[k\rho - \left(|\ell-\nu|+\frac{1}{2}\right)\frac{\pi}{2}\right].
\end{eqnarray}
The phase shift related to the scattering is therefore
\begin{equation}
\delta_{\ell}= -\frac{\pi}{2}|\ell-\nu|+\frac{\pi}{2}|\ell|.
\end{equation}
Observe that the phase shift vanishes in two cases, first when the defect disappears  ($b=0$) and, 
consequently, the electric field vanishes. In the second case when $\nu=0$, which 
implies in $b=\frac{\alpha \kappa B}{\hbar k}$. In both cases the quasiparticle is not scattered 
by the defect. In the first case, it is obvious that one does not obtain a scattering when the 
defect is absent. In second case, the combined effects of the electrical field of the defect 
and the external magnetic field compensates the effects of the stress and strain fields of the 
defects. 

The scattering amplitude for this system is
\begin{eqnarray}
f(\phi)=\sqrt{\frac{1}{2\pi i  k}}\{2\pi \delta(\phi - \pi)(1-\cos(\pi\nu))\nonumber\\-i\frac{\sin(\pi\nu)}{\cos(\frac{\phi}{2})}\exp\left[iN(\phi-\pi)-i\frac{\phi}{2}\right]\},
\end{eqnarray}
where  $N$ is the largest integer less than or equal to $\nu$. Note that we found a scattering 
behaviour similar to the Aharonov-Bohm effect, where the parameter $\nu$ represents a quantum 
potential. Notice that in \cite{bausch2} a chiral contribution to the scattering amplitude appears due to 
a repulsive term in the Hamiltonian derived from deformation potential theory. The fact that we did 
not consider neither the quadratic term in the electric field nor this repulsive term is quivalent to 
the case where we have both but they are adjusted to cancel each other (both terms are inversely 
proprtional to the radius square and have opposing signs).  

In summary, we have analyzed the quantum dynamics of a dipole induced in a neutral quasiparticle  
in the presence of a topological defect such as  a charged screw dislocation submitted  to an  
external magnetic  field parallel  to the defect. We showed that the scattering is a kind of 
Aharonov-Bohm  effect which can be found in solids. The analysis of the dipole dynamics allowed 
to determine the phase shift and the scattering amplitude. We also obtained the analytical 
expression for this amplitude and when we have the Burgers vector strength equal to 
$\alpha \kappa B/\hbar$, $\nu=0$ this implies that $\delta_{\ell}=0$ and $f(\phi)=0$. Therefore, 
we have no scattering. It means that we have the behaviour similar to the Ramsauer-Tousend effect 
of quantum mechanics,  where the scattering amplitude becomes zero. The scattering amplitude is of 
Aharonov-Bohm-type although an extra term, due to the deformation 
potential, not included in this treatment,  should appear \cite{bausch2}. As mentioned above a more detailed 
study, including this term,  is due in a forthcoming publication. The quasiparticle in the 
presence of this charged defect acquires  a topological quantum phase similar the phase 
experimented by  neutral atoms in the He-McKellar-Wilkens effect. We have shown that uncharged particles 
with an electric dipole moment moving in the presence of a charged screw dislocation acquire a quantum phase
which combines both the He-McKellar-Wilkens phase and   an additional phase due to the topology of the 
medium with defect. The problem of 
the quantum dynamics of an exciton  in the presence of a charged screw dislocation is then 
analogous to that of a quantum charged particle in the presence of a Aharonov-Bohm flux. The 
quantum phase experimented by an exciton is produced by a combined effect of the torsion of 
the defect, represented by its Burger's vector, by the electric field produced by  dangling 
bonds in the defect core and by the applied magnetic field. We hope that this effect can be 
observed experimentally in transport phenomena in materials with charged dislocations. In this 
letter we obtain the solid state analog of the He-McKellar-Wilkens effect for an exciton in the 
presence of charged screw dislocation.
 
\acknowledgments
This work was partially supported by CNPq, CAPES (PROCAD) and by PROINPE-UEFS.

\end{document}